\documentclass[twocol]{ametsocV6.1}
\title{Improvement of a neural network convection scheme by including triggering and evaluation in present and future climates}

\authors{Hugo Germain\aff{a}\correspondingauthor{Hugo Germain, hugo.germain@meteo.fr}, Blanka Balogh\aff{a}, Olivier Geoffroy\aff{a}, David Saint-Martin\aff{a}}

\affiliation{\aff{a}{Météo-France, CNRS, Univ. Toulouse, CNRM, Toulouse, France}}

\abstract{In this study, we improve a neural network (NN) parameterization of deep convection in the global atmosphere model ARP-GEM. To account for the sporadic nature of convection, we develop an NN parameterization that includes a triggering mechanism capable of detecting whether deep convection is active within a grid cell. This new data-driven parameterization outperforms the existing NN parameterization in present climate when replacing the original deep convection scheme of ARP-GEM. \textit{Online} simulations with the NN parameterization run without stability issues. Then, this NN parameterization is evaluated \textit{online} in a warmer climate. We confirm that using relative humidity instead of the specific total humidity as input for the NN (trained with present data) improves the performance and generalization in warmer climate. Finally, we perform the training of the NN parameterization with data from a warmer climate and this configuration get similar results when used in simulations in present or warmer climates.}

\begin{document}

\maketitle

\statement
This paper introduces a data-driven parameterization that significantly improves upon the method described in \cite{Balogh2025}. Two key advancements are presented, leading to reduced biases in the simulation using the data-driven parameterization. First, a triggering mechanism is incorporated in the data-driven parameterization, which effectively mitigates biases. Second, the replacement of absolute humidity with relative humidity as an input enhances both \textit{online} performance and stability, including in climates not encountered during the data-driven parameterization's training phase.

\section{Introduction}
Parameterizations of atmospheric moist processes are the main source of biases in current climate models \citep{medeiros_aquaplanets_2008, medeiros_revealing_2011, stevens_what_2013}. The use of Machine Learning (ML) techniques, especially Neural Networks (NNs), to develop data-driven parameterizations is a promising approach to significantly improve the accuracy of climate models \citep{gentine_could_2018}. During the past decade, data-driven approaches were widely used to develop parameterizations for climate models. NNs were used to produce accurate, yet numerically affordable radiative transfer schemes \citep[e.g.,][]{chevallier_neural_1998, krasnopolsky_new_2005, ukkonen_exploring_2022}, cloud microphysics \citep{sharma_superdropnet_2025, sarauer_physics-informed_2025} or convection \citep[e.g.,][]{brenowitz_interpreting_2020, Balogh2025}. They have been used to emulate subgrid-scale parameterizations from aggregated high-resolution simulations \citep[e.g.,][]{yuval_stable_2020, yuval_use_2021} or from a super-parameterized model \citep[e.g.,][]{gentine_could_2018, rasp_deep_2018}.

Until recent years, only a few simulations using data-driven parameterizations were carried out as a substitute for traditional physical ones. However, significant technical advancements in integrating NNs into Fortran-based models have now made it easier to perform \textit{online} tests of data-driven parameterizations. \cite{brenowitz_prognostic_2018} conducted an \textit{online} evaluation of a data-driven unified parameterization in a single column model, which was extended to a full General Circulation Model (GCM) in \cite{brenowitz_2019} and \cite{brenowitz_interpreting_2020}, with a focus on \textit{online} stability of the data-driven scheme. \cite{wang_stable_2022} also used NNs trained using SPCAM data to represent the subgrid-scale processes in the atmospheric model CAM5 \citep{neale_cam5_2012}. The NN parameterization described in \cite{watt-meyer_neural_2024} was based on the output of a global storm-resolving simulation using GFDL X-SHiELD \citep{harris_scientific_2021} to represent heating and moistening rates in the Global Forecast System \citep[GFS,][]{zhou_toward_2019}.  ClimSim Online \citep{yu_climSim-Online_2025} implemented Pytorch-Fortran \citep{alexeev_pytorch-fortran_2023} to conduct an experiment with a data-driven parameterization based on the ClimSim dataset \citep{yu_climSim_2023} in the E3SM model \citep{rasch_e3sm_2019}. Using FTorch \citep{atkinson_ftorch_2025} in the ICON-A model \citep{giorgetta_icon-A_2018}, several data-driven parameterizations were tested \textit{online}, such as deep convection \citep{heuer_interpretable_2024} (stable \textit{online} for 180 days) and radiative transfer \citep{hafner_stable_2025}. \cite{Balogh2025} (hereafter, B25) used the OASIS-coupler's Fortran/Python interface \citep{craig_development_2017} to replace the heating and moistening tendencies of a deep convection parameterization by NNs in the ARP-GEM global atmosphere model, version 1 \citep{Geoffroy_2025}. 

To evaluate the \textit{online} performance of the NN-based deep convection parameterization, B25 carried out a 30-year simulation using ARP-GEM. The simulation produced realistic physical fields for most variables. However, it exhibited some biases, particularly in high cloud cover and over the polar regions. In this paper, we aim to present two major improvements to the data-driven deep convection parameterization introduced in B25, addressing the biases we have identified using the ARP-GEM atmosphere model, version 2 \citep{geoffroy_global_2025}. The first improvement involves using a triggering mechanism. Second, following the suggestion in \cite{Beucler_2024}, we replace absolute humidity by relative humidity (RH) to improve the generalizability of the data-driven scheme. 

The following manuscript is organized as follows. The first section describes the data-driven parameterization, including the data-driven triggering mechanism, and its performance both \textit{offline} and \textit{online}. The second section extends the \textit{online} evaluation of the data-driven parameterization by testing its generalizability in a different climate. The last section contains the conclusion.

\section{An ML-parameterization with triggering mechanism}
\label{Sec2}

\subsection{Physical model description}
\label{SubSec2a}

We use the global, efficient and multi-resolution atmosphere model ARP-GEM version 2 \citep{geoffroy_global_2025} with minor modifications described below.  
The model configuration is the same as in B25 with a horizontal grid spacing of 55 km and 50 hybrid coordinate vertical levels, extending from the surface up to 2 hPa. The model time step is set to $\Delta t$ = 900 s.

Some modifications have been made to the model since the study of B25, hence our results are not directly comparable with it. B25 used ARP-GEM version 1, whereas here we use ARP-GEM version 2. The differences mainly concern the shallow convection scheme and model tuning. The triggering mechanism has also been slightly revised  with a modified formulation of entrainment in the test updraft used to determine whether the triggering criterion is met.
These differences are described in detail in \cite{geoffroy_global_2025}.

The deep convection parameterization of ARP-GEM is based on \cite{Tiedtke89} revised by \cite{Bechtold2008, Bechtold2014, IFSdoc, Geoffroy_2025, geoffroy_global_2025} and will be referred to as the Tiedtke-Bechtold scheme thereafter.
%Entrainment and detrainment rates are higher than in B25, with the coefficients $\epsilon_{up}$ and $\delta_{up}$, as defined in \cite{IFSdoc}, set to 2.0 $\cdot$ 10$^{-3}$ m$^{-1}$ and 0.8 $\cdot$ 10$^{-4}$ m$^{-1}$, respectively, instead of 1.8 $\cdot$ 10$^{-3}$ m$^{-1}$ and 0.75 $\cdot$ 10$^{-4}$ m$^{-1}$ in B25.
Additionally, the intensity of shallow convection is reduced by a factor of three. Finally, for simplicity, the shallow convection cloud cover is here set to zero instead of being parameterized. The differences in model physics, particularly those related to deep convection, explain the differences in results when replicating B25, as mentioned in Section \ref{Sec2}.\ref{SubSec2d}.

\subsection{The data-driven parameterizations}

The Tiedtke-Bechtold scheme computes atmospheric profiles of tendencies of dry static energy $\partial_t s$, specific humidity $\partial_t q$ and zonal and meridional winds. For simplicity, we only emulate the computations of thermodynamic tendencies ($\partial_t s$ and $\partial_t q$), given that they are the main tendencies of the deep convection scheme. The momentum tendencies are still computed by the Tiedtke-Bechtold parameterization. The deep convection parameterization that we seek to emulate is active only in the troposphere. Therefore, we have removed the top eight levels from each of the vertical profiles, describing the upper layers of the atmosphere. Hence, the output consists of two tendency profiles on 42 vertical levels each. Input and output variables and dimensions of the data-driven parameterizations are summarized in Table S1.

The simulation to generate the learning samples is a one-year AMIP-like simulation with forcings from the year 2005. The input and output variables are saved every three hours on the octahedral reduced Gaussian grid of ARP-GEM, so that each atmospheric column roughly covers the same area.

In the two following subsections, we introduce the two data-driven parameterizations used in this study: an NN parameterization designed and trained as in B25 and then the new parameterization addressing the limits of the B25 one.

\subsubsection{The B25 parameterization}

This first parameterization is defined as in B25. A Multi-Layer Perceptron predictor (referred to as MLP Predictor) with the same architecture as in B25 (six hidden layers of 1024 nodes each and activated by ReLU and an output layer with 84 nodes activated by a linear function) is trained using the Mean-Squared Error (MSE) loss to emulate the thermodynamic tendencies of the Tiedtke-Bechtold scheme.

This data-driven parameterization is trained using a similar dataset to that described in B25. From the outputs of the simulation generating the learning samples, we build a training dataset randomly selecting 20 000 columns (out of 136 000) at each saved model time step, yielding a learning sample of roughly 60 millions of columns. This first dataset will be noted, $\mathcal{D}_{B25}$.

\subsubsection{The parameterization with triggering}

Reproducing non-Gaussian processes can be challenging for NNs \citep{steininger_density-based_2021}, and they may generate artificial signals. This is particularly true for the representation of deep convection, given its episodic and threshold-dependent nature. Indeed, in our simulations, the Tiedtke-Bechtold scheme is not activated in about 90\% of the columns. However, the data-driven parameterization introduced in B25 produces deep convection that occurs in non-convective grid cells, adding background noise and leading to significant biases in areas where deep convection is uncommon, such as polar regions or in the high troposphere.

\begin{figure*}[!htbp]
    \centering
    \includegraphics[width=0.6\linewidth, clip, trim=4.7cm 9.4cm 4.7cm 8.9cm]{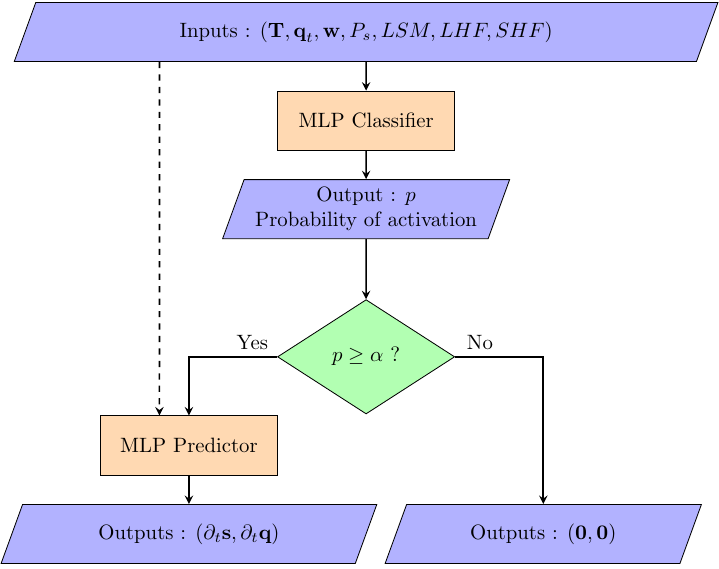}
    \caption{The parameterization with two NNs. Inputs : profiles (on 50 levels) of temperature ($\mathbf{T}$), specific total humidity ($\mathbf{q}_t$) and of vertical velocity ($\mathbf{w}$) and 4 scalar fields : land-sea mask (LSM), surface pressure ($P_s$), latent heat flux (LHF) and sensible heat flux (SHF). Outputs : profiles (on 42 levels) of dry static energy tendencies ($\partial_t\mathbf{s}$) and specific humidity tendencies ($\partial_t\mathbf{q}$). $\alpha$ is the minimal probability required for convection activation, it is a threshold to be tuned.}
    \label{Scheme2NNs}
\end{figure*}

To address this problem, we developed an NN parameterization that includes a triggering mechanism (Fig. \ref{Scheme2NNs}). The triggering mechanism is simply represented through a second neural network, a multilayer perceptron classifier (MLP), which is executed prior to the MLP Predictor, within the data-driven parameterization scheme. The MLP Classifier outputs the probability $p$ of deep convection activation within a grid cell given the same input as the MLP Predictor. If $p$ is greater than a threshold $\alpha$, the parameterization considers that the convection is active and the output tendencies are computed by the MLP Predictor. If not, the outputs are set to zero.

The MLP Classifier is composed of a total of seven layers : an input layer, five hidden layers of 1024, 1024, 512, 256 and 128 nodes respectively, and an output layer of a single node. All layers are activated by ReLU, expect for the last layer, which is activated by a sigmoid to yield values between 0 and 1. The MLP Classifier is trained using the binary cross-entropy loss function. The architecture and loss function of the MLP Predictor remains the same as in B25.

The learning dataset for this new parameterization is built using the outputs from the same 1-year ARP-GEM simulation, but we opt for a different subsampling strategy. Indeed, deep convection is active in only approximately 10\% of the atmospheric columns in the 2005 simulation. For this dataset, we retain all of the columns with active deep convection. The new learning sample is then completed by the addition randomly selected columns among the remaining 90\% of the simulated data, to build a learning sample with 50\% of the columns featuring deep convection. This second dataset will be noted $\mathcal{D}_{balanced}$. 

Both the MLP Classifier and the MLP Predictor are trained separately using the balanced dataset $\mathcal{D}_{balanced}$. Hereafter, the NN model resulting from this experiment will be noted as NN-t$\alpha$, where $\alpha$ represents the value of the triggering threshold applied in the MLP Classifier. For example, the NN using a threshold value of $\alpha=0.5$ will be noted, NN-t0.5. The NN-t0.0 configuration, for which the NN convection scheme is always active as in B25 (i.e. no triggering mechanism), differs from B25 parameterization only in its training dataset.

\begin{figure*}[!htbp]
    \centering
    \includegraphics[width=0.75\linewidth]{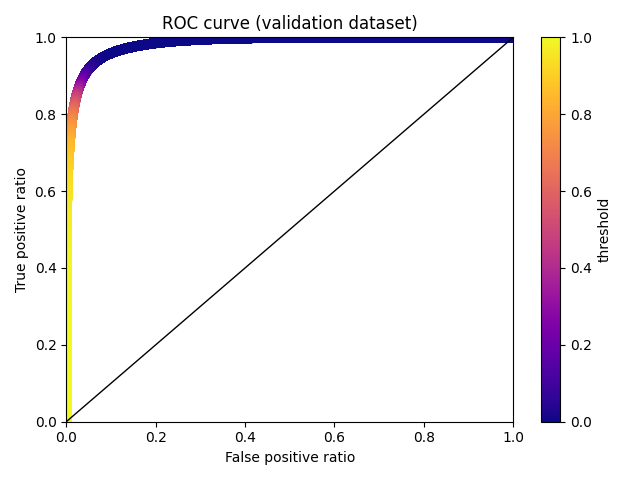}
    \caption{Receiver Operating Characteristic (ROC) curve of the MLP classifier.}
    \label{FigROC}
\end{figure*}

The threshold $\alpha$ of the triggering mechanism must be specified, after training. One way to do this is by using the Receiver Operating Characteristic (ROC) curve of the classifier (Fig. \ref{FigROC}). The MLP Classifier performs well: the curve almost reaches the point of coordinates (0, 1) (point which minimizes the false positive ratio while ensuring the highest possible true positive ratio), meaning that the MLP classifier effectively separates convectively active and inactive columns. For a first test, we chose the threshold $\alpha = 0.5$, which seems satisfying for the rest of the study. For this threshold, the proportion of active predicted columns is approximately 10\% (of the \textit{offline} test dataset), matching that of the true dataset.

\subsection{\textit{Offline} results}
\label{SubSec2c}

Once the training is achieved, we perform an \textit{offline} evaluation, conducted using data from another one-year-long AMIP simulation (2006). We have chosen a different year from the training dataset to that the training and validation datasets are independent. The outputs of Tiedtke-Bechtold scheme are considered as the reference values. In the 2006 simulation, the outputs are saved every twelve hours on a regular longitude-latitude grid to simplify the analysis of spatial patterns.

\begin{figure*}[!htbp]
    \centering
    \includegraphics[width=1\linewidth]{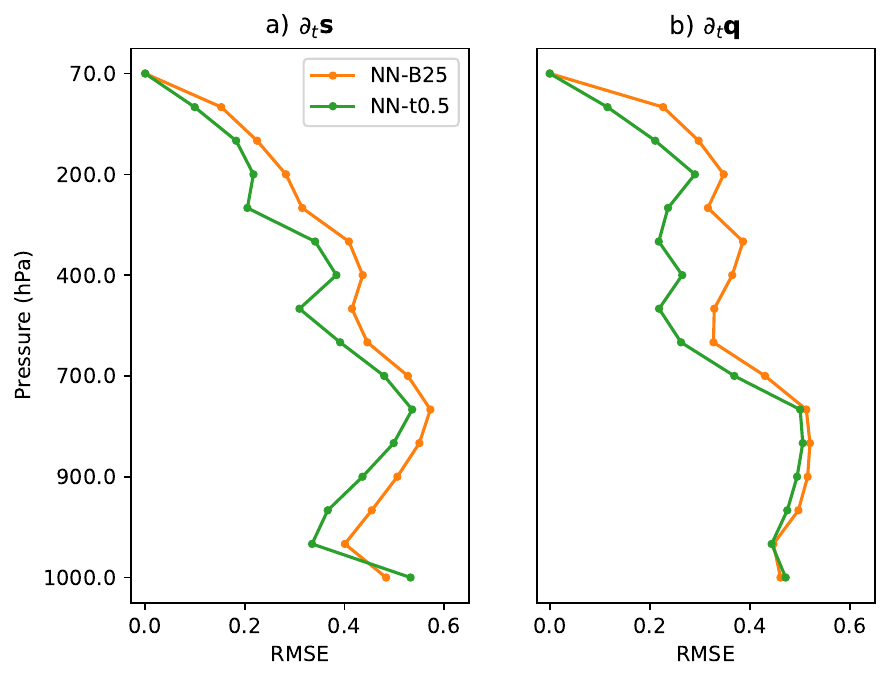}
    \caption{RMSE profiles of a) dry static energy and b) humidity tendencies for the NN-B25 parameterization, in orange and the NN-t0.5 parameterization, in green, computed on the validation dataset (year 2006) and interpolated to sixteen pressure levels.}
    \label{FigRMSE1NN2NN}
\end{figure*}

We compute the global root mean squared error (RMSE) of NN-t0.5 over the validation dataset. Fig. \ref{FigRMSE1NN2NN} shows the RMSE vertical profiles for the new and previous parameterizations, computed as follows:
\begin{equation}
%RMSE(k) = \sqrt{\frac{1}{N_tN_{lon}\sum_{i=1}^{N_{lat}}\cos{(\text{lat}_i)}} \sum_{n=1}^{N_t}\sum_{i=1}^{N_{lat}} \sum_{j=1}^{N_{lon}} \cos{(\text{lat}_i)}\left(y_{n,i,j,k} - y^{(NN)}_{n,i,j,k} \right)^2},
RMSE(k) = \sqrt{\frac{1}{N} \sum_{n=1}^{N_t}\sum_{i=1}^{N_{lat}} \sum_{j=1}^{N_{lon}} \cos{(\text{lat}_i)}\left(y_{n,i,j,k} - y^{(NN)}_{n,i,j,k} \right)^2},
\end{equation}
where $N = N_tN_{lon}\sum_{i=1}^{N_{lat}}\cos{(\text{lat}_i)}$. $N_t$, $N_{lat}, N_{lon}$ are the number timesteps, latitudes and longitudes, $\text{lat}_i$ the value of latitude for index $i$, $y_{n,i,j,k}$ the value of tendencies in the validation dataset and $y^{(NN)}_{n,i,j,k}$ the tendencies predicted by the NN, for the column with coordinates indexed by $(n,i,j,k)$.

At all levels except near the surface, the RMSE of the NN-t0.5 parameterization is lower than that of the NN-B25 parameterization, showing the benefits of the new sampling strategy for constructing the learning sample and the use of the data-driven triggering mechanism. 

The RMSE has low sensitivity to the triggering threshold $\alpha$. Only extreme values of $\alpha$ (0 or 1) result in significant changes in the score (Fig \ref{FigROC} and Fig. S2). The NN-t$\alpha$ parameterization yields better results than B25 across all thresholds in terms of global RMSE. The NN-t$\alpha$ parameterization yields the lowest RMSE when $\alpha$ = 0.0: no convective events are missed, which may play a role in the better \textit{offline} results obtained with $\alpha=0.0$. This setup enables us to isolate the effect of changes in the training dataset.

\begin{figure*}[!htbp]
    \includegraphics[width=\linewidth, clip, trim=2cm 7.3cm 1.5cm 7cm]{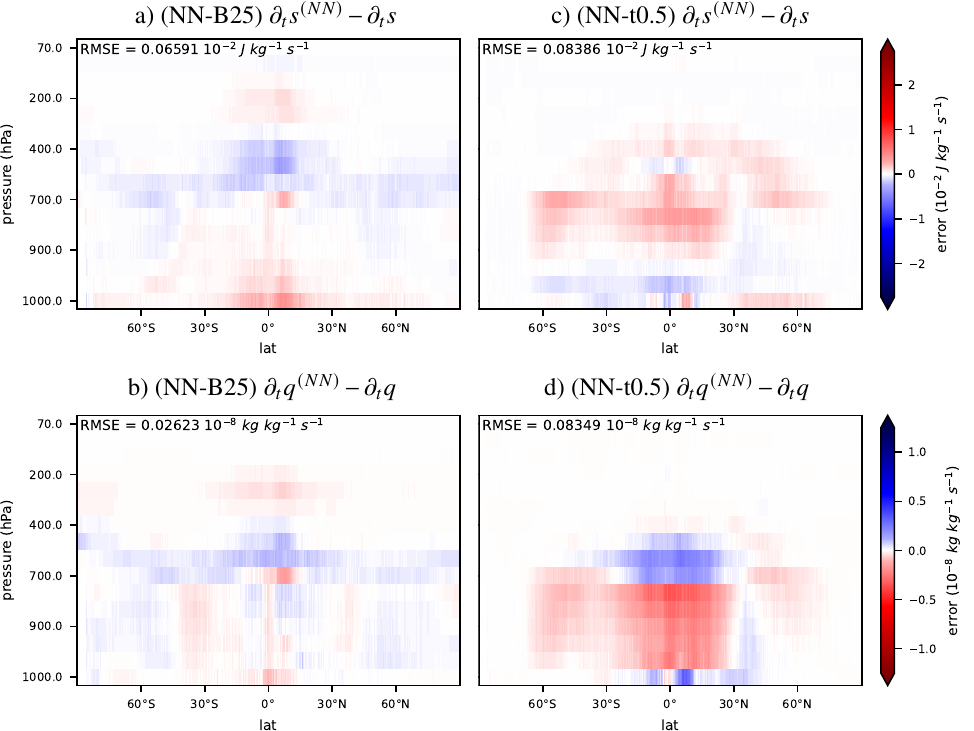}
    \caption{Differences from the zonal mean reference of a) dry static energy and b) humidity tendencies for the parameterization with 1 NN (NN-B25) and c) dry static energy and d) moisture tendencies for the NN-t0.5 parameterization computed on the validation dataset (year 2006). The RMSE values displayed are computed using Eq. \ref{eq:zonalRMSE}.}
    \label{FigZonal1NN2NN}
\end{figure*}

The raw zonal mean values are shown in Fig. S1: they show that both NN-B25 and NN-t0.5 are reproducing well the zonal means of the tendencies of the Tiedke-Bechtold scheme. Fig. \ref{FigZonal1NN2NN} shows the zonal mean differences between the NN and true tendencies. The displayed RMSE values are computed on the zonal means:
\begin{equation}
RMSE = \sqrt{\frac{1}{N'} \sum_{i=1}^{N_{lat}}\sum_{k=1}^{N_{lev}} \cos{(lat_i)}\left(\overline{y_{i,k}} - \overline{y_{i,k}^{(NN)}} \right)^2},
\label{eq:zonalRMSE}
\end{equation}
where $N' = N_{lev}\sum_{i=1}^{N_{lat}}\cos{(\text{lat}_i)}$. $N_{lev}$ is the number of vertical levels, $\overline{y}$ (resp. $\overline{y^{(NN)}}$) denotes the zonal mean of $y$ (resp. $y^{(NN)}$). We compare the zonal mean of the NN-B25 parameterization (Fig. \ref{FigZonal1NN2NN} a) and b)) with the NN-t0.5 parameterization (Fig. \ref{FigZonal1NN2NN} c) and d)). The NN-t0.5 parameterization has a higher RMSE value on zonal means than NN-B25. Indeed, the tendencies computed by NN-t0.5 are likely stronger because the training dataset $\mathcal{D}_{balanced}$ contains a proportionally larger number of columns exhibiting deep convection, leading to stronger predicted tendencies than those obtained with NN-B25. However, the anomalies at high latitudes (above 60°) disappear with the new NN architecture.

The zonal means of tendencies are more sensitive to the threshold than the RMSE computed for each column. For example, Fig. S3 (c and d) shows the zonal mean for the NN-t0.7 parameterization, which appear improved relative to Fig. \ref{FigZonal1NN2NN} (c and d). However, compared with an experiment where the physical deep convection scheme is not active (i.e., all tendencies are set to zero), the results are quite similar (not shown). In terms of zonal means, the NN-t0.0 parameterization shows very strong tendencies (Fig. S3 a and b), which mitigates its good per-column RMSE (Fig. S2). 

\textit{Offline} results may initially appear confusing, as they do not lead to a clear-cut conclusion. The NN-t0.5 configuration emerges as a good compromise, yielding an overall improvement in RMSE (Fig. \ref{FigRMSE1NN2NN} and Fig. S2). It effectively reduces background noise, especially at high latitudes, without significantly degrading the zonal means. The NN-t0.0 parameterization achieves the lowest RMSE but it produces zonal means that are too strong. These findings highlight the need for a comprehensive evaluation of the NN parameterizations through \textit{online} testing.

\subsection{\textit{Online} results}
\label{SubSec2d}
The next step in assessing the NN-t$\alpha$ parameterization is to use it \textit{online}, as a replacement for the thermodynamic tendencies produced by the physical parameterization of deep convection. Indeed, the \textit{offline} evaluation does not allow for an assessment of the performance of the data-driven parameterization in interaction with the rest of the physical model components, in particular the dynamical core. 

The numerical implementation of the data-driven parameterization follows the approach described in B25. ARP-GEM runs in parallel with a Python script responsible for executing the data-driven parameterization. Using the OASIS coupler \citep{craig_development_2017}, ARP-GEM sends the inputs to the Python script at each model time step. The Python script then executes the data-driven parameterization and returns the inferred heating and moistening tendencies to ARP-GEM. These tendencies are then added to the contributions from other physical parameterizations (e.g., turbulence, radiative transfer), and subsequently integrated by the dynamical core.

Four AMIP simulations are run over a five-year period (2006-2010), which is sufficient to exclude the contribution of interannual variability to differences between simulations. A description of the four simulations can be found in Table \ref{TabSim1}: ARP-GEM is the reference simulation using the Tiedtke-Bechtold parameterization; ARP-GEM (NN-B25), ARP-GEM (NN-t0.0) and ARP-GEM (NN-t0.5) are the three simulations in which it is replaced by the corresponding data-driven parameterization. We run the simulation ARP-GEM (NN-t0.0) to ensure that changes in the training dataset between B25 and NN-t0.5 are not the only source of differences.

\begin{table*}[!htbp]
\begin{center}
\begin{tabular}{lcccccc}
\topline
Reference Simulation Name & Climate &  &  &  &  & \\
\midline
ARP-GEM & Present \\
\midline
Simulation Name & Climate & NN type & Threshold $\alpha$ & Humidity variable & Training climate & Training dataset \\
\midline
ARP-GEM (NN-B25) & Present & NN-B25 & / & $q_t$ & Present & $\mathcal{D}_{B25}$ \\
ARP-GEM (NN-t0.0) & Present & NN-t$\alpha$ & 0.0 & $q_t$ & Present & $\mathcal{D}_{balanced}$ \\
ARP-GEM (NN-t0.5) & Present & NN-t$\alpha$ & 0.5 & $q_t$ & Present & $\mathcal{D}_{balanced}$ \\
\botline
\end{tabular}
\caption{Simulations analyzed in section \ref{Sec2}.\ref{SubSec2d}.}
\label{TabSim1}
\end{center}
\end{table*}

\begin{figure*}[!htbp]
    \centering
    \includegraphics[width=1\linewidth, clip, trim=2cm 5.7cm 1.8cm 2cm]{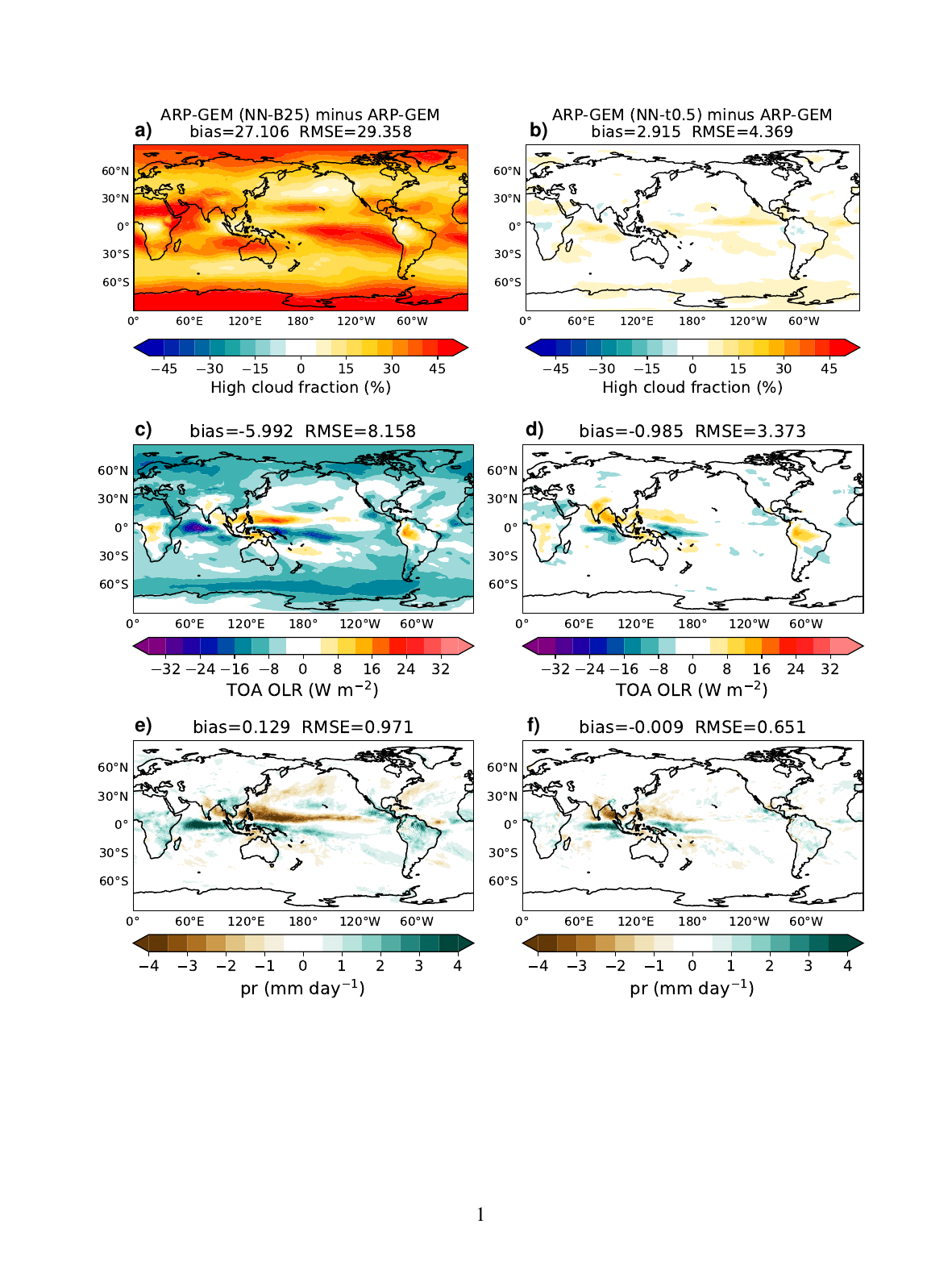}
    \caption{Anomaly with respect to an ARP-GEM reference simulation (with Tietdke-Bechtold scheme) for the simulation with the NN-B25 parameterization (a) high clouds, c) OLR, e) precipitations) and for the simulation with the NN-t0.5 parameterization (b) high clouds, d) OLR, f) precipitations).}
    \label{FigOnline1NN2NN}
\end{figure*}

For each simulation, we focus on key climate variables related to the radiation budget (namely, outgoing longwave radiation (OLR) and high cloud fraction) and precipitation. Fig. \ref{FigOnline1NN2NN}a) and b) show the anomaly of high cloud fraction with respect to the ARP-GEM reference simulation using NN-B25 (Fig \ref{FigOnline1NN2NN}a) and NN-t0.5 (Fig \ref{FigOnline1NN2NN}b). The anomaly for the NN-B25 parameterization is large, in particular at high latitudes and in the subsiding branch of the Hadley-Walker circulation in the tropics, such as over the eastern subtropical oceans. The increase in high cloud cover can be attributed to spurious convection events in these regions of small convective activity, leading to an excessive humidity detrainment rate (Fig. \ref{FigZonal1NN2NN}). In simulation ARP-GEM (NN-t0.5), the triggering mechanism enables the removal of this large bias and general reduction of error (Fig. \ref{FigOnline1NN2NN}b). Results from the ARP-GEM (NN-t0.0) simulation are shown in Fig. S4 (left). The simulation exhibits comparable errors in both high-level clouds and OLR, if not larger than those in NN-B25.
This shows that the change in the training dataset alone is not the cause of the bias reduction obtained with NN-t0.5.

With both NN-B25 and NN-t0.0, the positive bias in high cloud cover is associated with a large negative bias in OLR due to their enhanced greenhouse effect (Fig. \ref{FigOnline1NN2NN} c and Fig. S4). Consistently, these anomalies are strongly reduced when using the triggering mechanism (Fig. \ref{FigOnline1NN2NN}d). Note that the OLR and high cloud fraction biases were not as large in \cite{Balogh2025} (using ARP-GEM1) as those obtained with our version NN-B25 (using ARP-GEM2) (e.g., Fig. 4 in B25 and Fig. \ref{FigOnline1NN2NN}). These differences are related to updates in physics and tuning between the model versions used in each study. In particular, the deep convection tuning is different with more diluted updrafts in the present version, likely reaching lower levels. This may be the cause of the larger bias obtained with NN-B25 in comparison with \cite{Balogh2025}.

The precipitation field is strongly connected to deep convection which can bring a significant part of the annual precipitation amount, especially in the tropics \citep{nesbitt_storm_2006}. In addition, deep convection contributes to shaping the large-scale dynamics, which in turn influence the large-scale precipitation. Figures \ref{FigOnline1NN2NN}e and \ref{FigOnline1NN2NN}f show the precipitation anomaly with respect to the ARP-GEM reference simulation for e) the NN-B25 parameterization and f) the NN-t0.5 parameterization. For NN-B25, the main anomalies were located near the equator over the tropical Indian and Pacific Ocean and the warm pool. Again, the NN-t0.5 parameterization performs better than B25: the mean bias is close to zero and the RMSE is significantly reduced. The spatial anomalies also appear weaker and are concentrated over the Maritime Continent, the eastern of tropical Indian Ocean and the western tropical Pacific Ocean. The ARP-GEM (NN-t0.0) experiment exhibits a mean precipitation field (Fig. S4 right) comparable to those of ARP-GEM (NN-t0.5), with both outperforming the results obtained using NN-B25. Thus, since NN-t0.5 performs better across all variables (not only precipitation), the main source of improvement over NN-B25 lies in the combined use of the triggering mechanism and the use of the balanced training dataset. 

\begin{figure*}[!htbp]
    \centering
    \includegraphics[width=\linewidth, clip, trim=2cm 6.5cm 2cm 4.1cm]{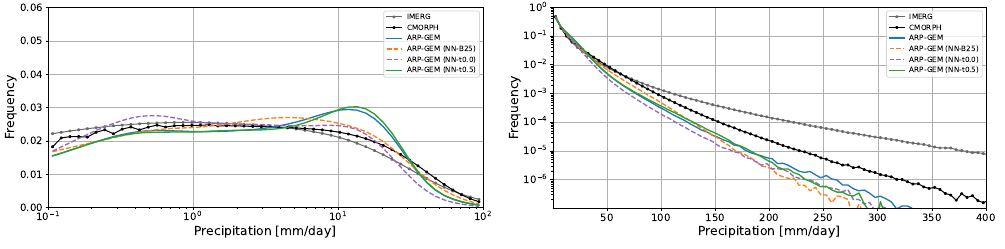}
    \caption{Probability density functions (PDFs) of daily precipitations for observational datasets interpolated to a regular 0.5° grid (IMERG in grey and CMORPH in black), an ARP-GEM reference simulation with Tiedtke-Bechtold parameterization (ARP-GEM in blue), an ARP-GEM simulation with the NN-B25 parameterization (ARP-GEM (NN-B25) in orange), an ARP-GEM simulation with the NN-t0.0 parameterization (ARP-GEM (NN-t0.0) in purple) and an ARP-GEM simulation with the NN-t0.5 parameterization (ARP-GEM (NN-t0.5) in green). PDFs are computed only for latitudes between 20°N and 20°S.}
    \label{FigPDF1NN2NN}
\end{figure*}

The use of NN-t0.5 leads to significantly improved radiative fields compared with NN-B25. For precipitation, biases can be compared against the observational climatology (Fig. S5). The bias of ARP-GEM relative to the climatology is approximately 1 mm day$^{-1}$ and the bias reduction from ARP-GEM (NN-B25) to ARP-GEM (NN-t0.5) is approximately 0.3 mm day$^{-1}$, corresponding to a 30\% reduction of ARP-GEM bias, which is significant.
%For all the other variables (shortwave (SW) radiation, other cloud layers) the NN-t0.5 parameterization outperforms the NN-B25 one (not shown).
To assess variability, we compute the probability density functions (PDFs) of daily precipitations for the observational datasets IMERG \citep{huffman_2019}, CMORPH \citep{xie_2017} and the four ARP-GEM simulations (Fig. \ref{FigPDF1NN2NN}). The experiment using the data-driven parameterization with triggering (NN-t0.5) is closer to the reference ARP-GEM simulation than those using NN-B25 and NN-t0.0.

The impact of the threshold $\alpha$ in \textit{online} performance was investigated, too. Similar to the \textit{offline} evaluation, the sensitivity to the value of $\alpha$ in the \textit{online} experiments is also low. As expected, the performance drops only when $\alpha$ is set to 0 (i.e. no triggering) or 1 (i.e. no parameterized deep convection). For other values (typically between 0.1 and 0.9) the performances remains approximately the same (not shown).

Finally, the new data-driven parameterization NN-t0.5 outperforms that introduced in B25, showing significant improvements in both the mean fields and the representation of variability. This validates the choice of an additional data-driven triggering mechanism, inspired by the Tiedtke-Bechtold scheme, along with an improved sampling strategy for building the training dataset. Therefore, in the following sections, only NN-t0.5 will be used.  %Consequently, the data-driven parameterization including the triggering mechanism and the use of the $\mathcal{D}_{balanced}$ dataset to train the parameterization is retained for the remainder of the study.

\section{Evaluation in warmer climate}
\label{Sec3}

NN parameterizations often demonstrate limited extrapolation capabilities beyond the training data distribution. In climate modeling, this can happen when NNs are applied to climates that differ from those sampled during training. When using data-driven parameterizations, this could lead to stability issues \citep{brenowitz_2019, brenowitz_2020} and degraded performances \citep{ogorman_2018}. 
In this section, we aim to study the generalizability of the data-driven parameterization in climates not sampled during training. First, we evaluate the performance in future (warmer) climate of the NN parameterization trained on present climate and investigating the impact of using relative humidity as an input instead of absolute humidity. Then, we extend the study to an NN trained on data sampled in warmer climate instead of current climate. 

\subsection{Choice of humidity predictor: $q$ vs. $RH$}
\label{SubSec3a}

In order to get an \textit{offline} validation dataset of a warmer climate, we run one year of simulation (year 2006) for which the prescribed sea surface temperature forcing is increased by 4K \citep{cess_methodology_1988, bony_cfmip_2011}. We will call this climate, +4K climate. To mitigate potential extrapolation issues, we aim to use variables with consistent value ranges across both current and +4K climates. Relative Humidity ($RH$) provides a clear example: its distribution remains consistent across climates \citep{Manabe1967}, and using $RH$ as the humidity predictor leads to improved (\textit{offline}) performance of the data-driven parameterization compared with $q_t$ \citep{Beucler_2024}. To assess the impact of the choice of the humidity variable on the extrapolation capabilities of the data driven model, we trained both NNs of the parameterization using $RH$ instead of $q_t$ using data from a simulation in current climate ($\mathcal{D}_{balanced}$).

When evaluated in current climate, the \textit{offline} performance of the NN parameterization remains unchanged, regardless of whether $q_t$ or $RH$ is used as the humidity input (not shown). In the +4K climate, the parameterization using $RH$ performs better at nearly all levels, despite an overall degradation in performance compared to results in current climate (Fig. S6). This degradation can be attributed to other input variables, whose value ranges vary across climate. Zonal mean differences further support that the use of RH instead of $q_t$ improves performance (not shown).

To evaluate \textit{online} performance of the data-driven parameterizations in +4K, we run three five-year experiments in a +4K climate, described in Table \ref{TabSim2} : ARP-GEM+4K, the reference simulation as well as ARP-GEM+4K (NN-t0.5-$q_t$) and ARP-GEM+4K (NN-t0.5-RH), the +4K simulation where the deep convection scheme is replaced by the data-driven parameterizations using $q_t$ and $RH$ as humidity variables, respectively.

\begin{table*}[!htbp]
\begin{center}
\begin{tabular}{lcccccc}
\topline
Reference Simulation Name & Climate &  &  &  &  & \\
\midline
ARP-GEM+4K & +4K \\
\midline
Simulation Name & Climate & NN type & Threshold & Humidity variable & Training climate & Training dataset \\
\midline
ARP-GEM+4K (NN-t0.5-$q_t$) & +4K & NN-t$\alpha$ & 0.5 & $q_t$ & Present & $\mathcal{D}_{balanced}$ \\
ARP-GEM+4K (NN-t0.5-RH) & +4K & NN-t$\alpha$ & 0.5 & RH & Present & $\mathcal{D}_{balanced}$ \\
\botline
\end{tabular}
\caption{Simulations analyzed in section \ref{Sec3}.\ref{SubSec3a}.}
\label{TabSim2}
\end{center}
\end{table*}

\begin{figure*}[!htbp]
    \centering
    \includegraphics[width=1\linewidth, clip, trim=0cm 0.6cm 0cm 10cm]{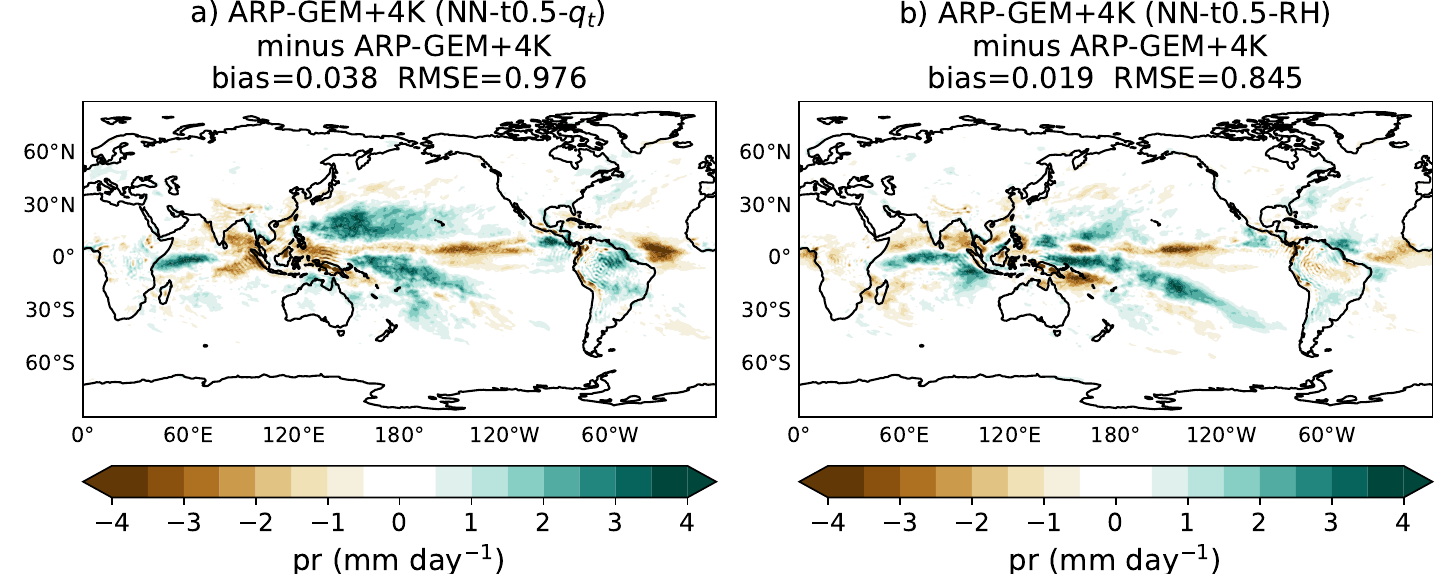}
    \caption{Precipitation anomaly with respect to an ARP-GEM+4K reference simulation (with Tietdke-Bechtold scheme) in a +4K climate for a) the simulation with the NN-t0.5-$q_t$ parameterization and b) the simulation with the NN-t0.5-RH parameterization.}
    \label{FigQTRH4Kpr}
\end{figure*}

Both simulations using data-driven parameterizations remain stable for five years, regardless of the choice of humidity variable. This indicates that the extrapolation capabilities of the data-driven parameterizations are sufficient to maintain the stability of the simulation. 
%We compared the results of ARP-GEM+4K (NN-t0.5-$q_t$) and ARP-GEM+4K (NN-t0.5-RH) with respect to the reference simulation in a +4K climate (we do not look at climate change tendencies). 
Fig. \ref{FigQTRH4Kpr} shows the anomalies with respect to ARP-GEM+4K in terms of precipitations of ARP-GEM+4K (NN-t0.5-$q_t$) and ARP-GEM+4K (NN-t0.5-$RH$). Although the results, both in terms of pattern and intensity, are degraded relative to the \textit{online} validation in current climate (RMSE increases from 0.651 to 0.976 mm day$^{-1}$ for $q_t$), using the parameterization trained with $RH$ as the humidity predictor leads to improved \textit{online} performances (RMSE = 0.845 mm day$^{-1}$).

The parameterization using RH instead of $q_t$ tends to perform better for precipitation (Fig.\ref{FigQTRH4Kpr}) and OLR ($RMSE_{q_t} = 4.98\ Wm^{-2}$ and $RMSE_{RH} = 3.72\ Wm^{-2}$) during the \textit{online} evaluation in the +4K climate. However, for top of atmosphere shortwave (SW) radiation, the parameterization using $q_t$ more accurately reproduces the mean field pattern (Fig. S7). The larger errors with relative humidity may be related to an excess of cloud liquid water, though this aspect is not investigated further. Since using $RH$ results in a general improvement of performance of the data-driven parameterization, we adopt this variable for the remainder of the study.

\subsection{Training in +4K climate}
\label{Subsec3b}

To construct a training dataset for the +4K climate, we proceed following the same method as for current climate, i.e., using a one-year ARP-GEM simulation (2005) but with the prescribed sea surface temperatures increased by 4 K. We then test an NN parameterization trained on +4K climate data (denoted \textit{NN +4K}) and compare it to the NN parameterization trained on current climate data (denoted \textit{NN present}) in both current and future climates.

\textit{Offline} and \textit{online} validation lead to the same conclusions. Therefore, we focus on the \textit{online} results. We analyze four simulations, all of which remain stable over five years, and compare them with the reference simulations. The experiences are described in Table \ref{TabSim3}. Note that since all \textit{online} simulations use the NN-t0.5-$RH$ configuration, the simulation names in Table \ref{TabSim3} have been simplified.

\begin{table*}[!htbp]
\begin{center}
\begin{tabular}{lcccccc}
\topline
Reference Simulation Name & Climate &  &  &  &  & \\
\midline
ARP-GEM & Present \\
ARP-GEM+4K & +4K \\
\midline
Simulation Name & Climate & NN type & Threshold & Humidity variable & Training climate & Training dataset \\
\midline
ARP-GEM (NN present) & Present & NN-t$\alpha$ & 0.5 & RH & Present & $\mathcal{D}_{balanced}$ \\
ARP-GEM (NN +4K) & Present & NN-t$\alpha$ & 0.5 & RH & +4K & $\mathcal{D}_{balanced}$ \\
ARP-GEM+4K (NN present) & +4K & NN-t$\alpha$ & 0.5 & RH & Present & $\mathcal{D}_{balanced}$ \\
ARP-GEM+4K (NN +4K) & +4K & NN-t$\alpha$ & 0.5 & RH & +4K & $\mathcal{D}_{balanced}$ \\
\botline
\end{tabular}
\caption{Simulations analyzed in section \ref{Sec3}.\ref{Subsec3b}.}
\label{TabSim3}
\end{center}
\end{table*}

\begin{figure*}[!htbp]
    \centering
    \includegraphics[width=1\linewidth, clip, trim=2.5cm 12.7cm 2.2cm 2.2cm, scale=0.5]{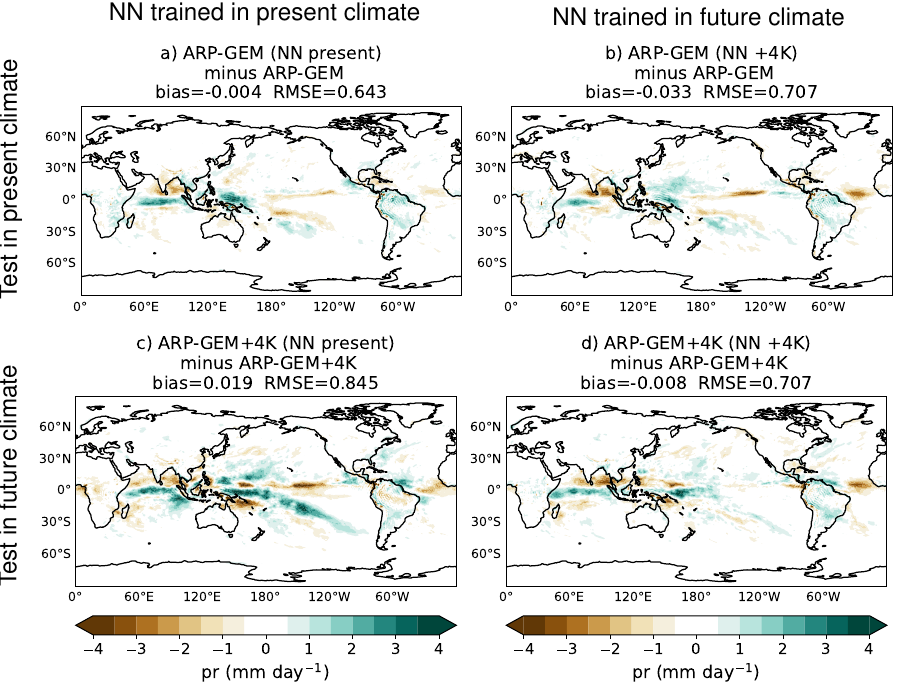}
    \caption{Precipitation anomaly in present climate with respect to a present ARP-GEM reference simulation for the parameterization trained with a) present data and b) +4K data tested in present and precipitation anomaly in +4K climate with respect to a +4K ARP-GEM reference simulation for the parameterization trained with c) present data and d) +4K data tested in +4K.}
    \label{FigOnlineMatrix}
\end{figure*}

Fig. \ref{FigOnlineMatrix} shows online results (in present and +4K climates) of the NN parameterizations trained on present climate data (Fig. \ref{FigOnlineMatrix} a) and c) (left column)) compared to the data driven parameterization trained on +4K climate data (Fig. \ref{FigOnlineMatrix} b) and d) (right column)). The results presented in the first column of this figure correspond to those discussed in Section \ref{Sec2}\ref{SubSec2d} and Section \ref{Sec3}\ref{SubSec3a}: the data-driven parameterization trained using present climate data exhibits reduced accuracy when used in a +4K simulation compared to its performance in a current climate experiment. Conversely, the NN parameterization trained on +4K climate data performs effectively in the +4K scenario (Fig. \ref{FigOnlineMatrix} d), although it remains slightly less accurate than an NN trained and tested exclusively in current climate (Fig. \ref{FigOnlineMatrix} a). Notably, the NN trained on +4K data also outperforms, when used in a current climate simulation (Fig. \ref{FigOnlineMatrix} b), an NN trained on present-day data but tested in a +4K scenario (Fig. \ref{FigOnlineMatrix} c). These findings, which are valid for other variables and \textit{offline} experiments (not shown), are consistent with the results reported by \cite{ogorman_2018}. Their work demonstrated that extra-tropical atmospheric columns in a +4K climate provide relevant information to understand tropical columns in the present climate.

\section{Conclusion}

This study aims at improving an NN parameterization of deep convection in a climate model, namely, ARP-GEM at 55 km horizontal resolution. We found that incorporating physical knowledge (the triggering mechanism or the use of $RH$ instead of $q_t$) in the development of data-driven parameterizations could lead to more accurate results. First, we introduced an NN parameterization with a modified training dataset and mainly a triggering mechanism that can detect the activation of convection. This new architecture enables us to correct small errors that were occurring in grid cells where the deep convection scheme is not supposed to be active and were amplifying in \textit{online} tests.

This parameterization effectively distinguishes the cases when deep convection is active or from those when it is not. In particular, \textit{offline} evaluation shows that the NN parameterization with triggering outperforms those described in B25 in terms of global RMSE, although its zonal mean performance is slightly degraded. When implemented in ARP-GEM, the NN parameterization with triggering clearly outperforms the B25 parameterization in present climate. In particular, the parameterization with the triggering mechanism significantly reduces \textit{online} biases, especially for high clouds and OLR, compared with the basic NN parameterization. Improvements are also observed in the representation of the precipitation with  the daily precipitation PDF being particularly well represented with the triggering. The threshold $\alpha$ introduced for the triggering mechanism has a limited impact on the outputs of the NN parameterization, except at extreme values (0 or 1). This type of parameterization, which includes a triggering mechanism that prevents the generation of noisy values instead of exact zeros, could be used in other applications, in particular when an NN is used to emulate processes that occur intermittently.

We then tested the NN parameterization with triggering in a warmer (+4K) climate, using either of two different humidity variables as input ($RH$ and $q_t$). The simulation using a data-driven parameterization trained only on present climate data remains stable over five years with either humidity variable. Performance is slightly degraded compared with the results in present climate. Outside of the training climate, using $RH$ instead of $q_t$ as input improved the out-of-sample generalization of the parameterization.

Finally, we trained an NN parameterization using data from a +4K simulation, which produced accurate results when tested on data from the warmer climate. Unlike the parameterization trained on current climate, which was less accurate in the +4K experiment, the parameterization trained on +4K data generalizes well in the current (colder) climate experiment. These findings are consistent with \cite{ogorman_2018}.

When replacing deep convection only with NNs, we did not encounter stability issues. However, this does not guarantee \textit{online} stability if additional or different data-driven components replace other physical parameterizations. The next step is to move beyond emulating existing physical parameterizations and use aggregated output from reanalysis and/or kilometer-scale climate simulations to train data-driven components to enhance climate simulations. 

% Acknowledgments & datastatements are needed at the end of the paper
\clearpage

\acknowledgments
%  Keep acknowledgments (note correct spelling: no ``e'' between the ``g'' and
% ``m'') as brief as possible. In general, acknowledge only direct help in
%  writing or research. Financial support (e.g., grant numbers) for the work done, 
%  for an author, or for the laboratory where the work was performed must be 
%  acknowledged here rather than as footnotes to the title or to an author's name.
%  Contribution numbers (if the work has been published by the author's institution 
%  or organization) should be placed in the acknowledgments rather than as 
%  footnotes to the title or to an author's name.

%%%%%%%%%%%%%%%%%%%%%%%%%%%%%%%%%%%%%%%%%%%%%%%%%%%%%%%%%%%%%%%%%%%%%
% DATA AVAILABILITY STATEMENT
%%%%%%%%%%%%%%%%%%%%%%%%%%%%%%%%%%%%%%%%%%%%%%%%%%%%%%%%%%%%%%%%%%%%%
% 
%
\datastatement
The supporting dataset and code are available on Zenodo: https://doi.org/10.5281/zenodo.17531476.

%%%%%%%%%%%%%%%%%%%%%%%%%%%%%%%%%%%%%%%%%%%%%%%%%%%%%%%%%%%%%%%%%%%%%
% REFERENCES
%%%%%%%%%%%%%%%%%%%%%%%%%%%%%%%%%%%%%%%%%%%%%%%%%%%%%%%%%%%%%%%%%%%%%
\bibliographystyle{ametsocV6}
\bibliography{references}

@STRING{AN        = "Astrophys.\ Norv."}

@STRING{MA        = "Meteor.\ Appl."}

@article {Balogh2025,
      author = "Blanka Balogh and David Saint-Martin and Olivier Geoffroy",
      title = "Online Test of a Neural Network Deep Convection Parameterization in {ARP-GEM}1",
      journal = "Artificial Intelligence for the Earth Systems",
      year = "2025",
      publisher = "American Meteorological Society",
      address = "Boston MA, USA",
      volume = "4",
      number = "3",
      doi = "10.1175/AIES-D-24-0100.1",
      pages=      "240100",
      url = "https://journals.ametsoc.org/view/journals/aies/4/3/AIES-D-24-0100.1.xml"
}

@article {Tiedtke89,
      author = "M.  Tiedtke",
      title = "{A Comprehensive Mass Flux Scheme for Cumulus Parameterization in Large-Scale Models}",
      journal = "Monthly Weather Review",
      year = "1989",
      publisher = "American Meteorological Society",
      address = "Boston MA, USA",
      volume = "117",
      number = "8",
      doi = "10.1175/1520-0493(1989)117%3C1779:ACMFSF%3E2.0.CO;2",
      pages=      "1779 - 1800",
      url = "https://journals.ametsoc.org/view/journals/mwre/117/8/1520-0493_1989_117_1779_acmfsf_2_0_co_2.xml"
}

@article{Bechtold2008,
    author = {Bechtold, Peter and Köhler, Martin and Jung, Thomas and Doblas-Reyes, Francisco and Leutbecher, Martin and Rodwell, Mark J. and Vitart, Frederic and Balsamo, Gianpaolo},
    title = "{Advances in simulating atmospheric variability with the ECMWF model: From synoptic to decadal time-scales}",
    journal = {Quarterly Journal of the Royal Meteorological Society},
    volume = {134},
    number = {634},
    pages = {1337-1351},
    doi = {10.1002/qj.289},
    year = {2008}
}

@article {Bechtold2014,
      author = "Peter Bechtold and Noureddine Semane and Philippe Lopez and Jean-Pierre Chaboureau and Anton Beljaars and Niels Bormann",
      title = "{Representing Equilibrium and Nonequilibrium Convection in Large-Scale Models}",
      journal = "Journal of the Atmospheric Sciences",
      year = "2014",
      publisher = "American Meteorological Society",
      address = "Boston MA, USA",
      volume = "71",
      number = "2",
      doi = "10.1175/JAS-D-13-0163.1",
      pages=      "734 - 753",
      url = "https://journals.ametsoc.org/view/journals/atsc/71/2/jas-d-13-0163.1.xml"
}

@inbook{IFSdoc,
  author = {ECMWF},
  title = {IFS Documentation CY49R1 - Part IV: Physical Processes},
  year = {2024},
  journal = {IFS Documentation CY49R1},
  chapter = {4},
  month = {11/2024},
  publisher = {ECMWF},
  doi = {10.21957/c731ee1102},
}

@article{Geoffroy_2025,
	title = {The {ARP}-{GEM1} {Global} {Atmosphere} {Model}: {Description}, {Speedup} {Analysis}, and {Multiscale} {Evaluation} up to 6 km},
	volume = {38},
	issn = {0894-8755, 1520-0442},
	shorttitle = {The {ARP}-{GEM1} {Global} {Atmosphere} {Model}},
	url = {https://journals.ametsoc.org/view/journals/clim/38/18/JCLI-D-24-0547.1.xml},
	doi = {10.1175/JCLI-D-24-0547.1},
	language = {EN},
	number = {18},
	urldate = {2025-09-08},
	journal = {Journal of Climate},
	author = {Geoffroy, Olivier and Saint-Martin, David},
	month = aug,
	year = {2025},
	note = {Publisher: American Meteorological Society Section: Journal of Climate},
	pages = {4739--4762},
}

@article{hafner_stable_2025,
title={Stable Machine Learning based Radiation Emulation for ICON},
url={http://dx.doi.org/10.22541/essoar.174708082.27787580/v1},
DOI={10.22541/essoar.174708082.27787580/v1},
publisher={Wiley},
author={Hafner, Katharina and Iglesias-Suarez, Fernando and Shamekh, Sara and Gentine, Pierre and Giorgetta, Marco A. and Pincus, Robert and Eyring, Veronika},
year={2025},
month=may }

@article{heuer_interpretable_2024,
author = {Heuer, Helge and Schwabe, Mierk and Gentine, Pierre and Giorgetta, Marco A. and Eyring, Veronika},
title = {Interpretable Multiscale Machine Learning-Based Parameterizations of Convection for ICON},
journal = {Journal of Advances in Modeling Earth Systems},
volume = {16},
number = {8},
pages = {e2024MS004398},
doi = {10.1029/2024MS004398},
url = {https://agupubs.onlinelibrary.wiley.com/doi/abs/10.1029/2024MS004398},
note = {e2024MS004398 2024MS004398},
year = {2024}
}

@article{rasp_deep_2018,
author = {Stephan Rasp  and Michael S. Pritchard  and Pierre Gentine },
title = {Deep learning to represent subgrid processes in climate models},
journal = {Proceedings of the National Academy of Sciences},
volume = {115},
number = {39},
pages = {9684-9689},
year = {2018},
doi = {10.1073/pnas.1810286115},
URL = {https://www.pnas.org/doi/abs/10.1073/pnas.1810286115},
}

@article{wang_stable_2022,
AUTHOR = {Wang, X. and Han, Y. and Xue, W. and Yang, G. and Zhang, G. J.},
TITLE = {Stable climate simulations using a realistic general circulation model with neural network
parameterizations for atmospheric moist physics and radiation processes},
JOURNAL = {Geoscientific Model Development},
VOLUME = {15},
YEAR = {2022},
NUMBER = {9},
PAGES = {3923--3940},
URL = {https://gmd.copernicus.org/articles/15/3923/2022/},
DOI = {10.5194/gmd-15-3923-2022}
}

@article{brenowitz_2019,
	title = {Spatially {Extended} {Tests} of a {Neural} {Network} {Parametrization} {Trained} by {Coarse}-{Graining}},
	volume = {11},
	copyright = {©2019. The Authors.},
	issn = {1942-2466},
	url = {https://onlinelibrary.wiley.com/doi/abs/10.1029/2019MS001711},
	doi = {10.1029/2019MS001711},
	language = {en},
	number = {8},
	urldate = {2025-07-18},
	journal = {Journal of Advances in Modeling Earth Systems},
	author = {Brenowitz, Noah D. and Bretherton, Christopher S.},
	year = {2019},
	pages = {2728--2744},
}

@misc{brenowitz_2020,
	title = {Machine {Learning} {Climate} {Model} {Dynamics}: {Offline} versus {Online} {Performance}},
	shorttitle = {Machine {Learning} {Climate} {Model} {Dynamics}},
	doi = {10.48550/arXiv.2011.03081},
	urldate = {2025-09-08},
	publisher = {arXiv},
	author = {Brenowitz, Noah D. and Henn, Brian and McGibbon, Jeremy and Clark, Spencer K. and Kwa, Anna and Perkins, W. Andre and Watt-Meyer, Oliver and Bretherton, Christopher S.},
	month = nov,
	year = {2020},
}

@article{watt-meyer_neural_2024,
author = {Watt-Meyer, Oliver and Brenowitz, Noah D. and Clark, Spencer K. and Henn, Brian and Kwa, Anna and McGibbon, Jeremy and Perkins, W. Andre and Harris, Lucas and Bretherton, Christopher S.},
title = {Neural Network Parameterization of Subgrid-Scale Physics From a Realistic Geography Global Storm-Resolving Simulation},
journal = {Journal of Advances in Modeling Earth Systems},
volume = {16},
number = {2},
pages = {e2023MS003668},
doi = {10.1029/2023MS003668},
note = {e2023MS003668 2023MS003668},
year = {2024}
}

@Article{medeiros_revealing_2011,
author={Medeiros, Brian
and Stevens, Bjorn},
title={Revealing differences in GCM representations of low clouds},
journal={Climate Dynamics},
year={2011},
month={Jan},
day={01},
volume={36},
number={1},
pages={385-399},
issn={1432-0894},
doi={10.1007/s00382-009-0694-5},
url={https://doi.org/10.1007/s00382-009-0694-5}
}

@article{stevens_what_2013,
author = {Bjorn Stevens  and Sandrine Bony },
title = {What Are Climate Models Missing?},
journal = {Science},
volume = {340},
number = {6136},
pages = {1053-1054},
year = {2013},
doi = {10.1126/science.1237554},
URL = {https://www.science.org/doi/abs/10.1126/science.1237554},
}

@article {medeiros_aquaplanets_2008,
author = "Brian Medeiros and Bjorn Stevens and Isaac M. Held and Ming Zhao and David L. Williamson and Jerry G. Olson and Christopher S. Bretherton",
title = "Aquaplanets, Climate Sensitivity, and Low Clouds",
journal = "Journal of Climate",
year = "2008",
publisher = "American Meteorological Society",
address = "Boston MA, USA",
volume = "21",
number = "19",
doi = "10.1175/2008JCLI1995.1",
pages=      "4974 - 4991",
url = "https://journals.ametsoc.org/view/journals/clim/21/19/2008jcli1995.1.xml"
}

@article{gentine_could_2018,
author = {Gentine, P. and Pritchard, M. and Rasp, S. and Reinaudi, G. and Yacalis, G.},
title = {Could Machine Learning Break the Convection Parameterization Deadlock?},
journal = {Geophysical Research Letters},
volume = {45},
number = {11},
pages = {5742-5751},
doi = {10.1029/2018GL078202},
url = {https://agupubs.onlinelibrary.wiley.com/doi/abs/10.1029/2018GL078202},
year = {2018}
}

@article{sharma_superdropnet_2025,
author = {Sharma, Shivani and Greenberg, David S.},
title = {SuperdropNet: A Stable and Accurate Machine Learning Proxy for Droplet-Based Cloud Microphysics},
journal = {Journal of Advances in Modeling Earth Systems},
volume = {17},
number = {6},
pages = {e2024MS004279},
doi = {10.1029/2024MS004279},
url = {https://agupubs.onlinelibrary.wiley.com/doi/abs/10.1029/2024MS004279},
note = {e2024MS004279 2024MS004279},
year = {2025}
}

@article{sarauer_physics-informed_2025, 
title={A physics-informed machine learning parameterization for cloud microphysics in ICON},
volume={4}, 
DOI={10.1017/eds.2025.10016}, 
journal={Environmental Data Science}, 
author={Sarauer, Ellen and Schwabe, Mierk and Weiss, Philipp and Lauer, Axel and Stier, Philip and Eyring, Veronika}, 
year={2025}, 
pages={e40}
}

@article{yuval_stable_2020,
author={Yuval, Janni
and O'Gorman, Paul A.},
title={Stable machine-learning parameterization of subgrid processes for climate modeling at a range of resolutions},
journal={Nature Communications},
year={2020},
month={Jul},
day={03},
volume={11},
number={1},
pages={3295},
issn={2041-1723},
doi={10.1038/s41467-020-17142-3},
url={https://doi.org/10.1038/s41467-020-17142-3}
}

@article{yuval_use_2021,
author = {Yuval, Janni and O'Gorman, Paul A. and Hill, Chris N.},
title = {Use of Neural Networks for Stable, Accurate and Physically Consistent Parameterization of Subgrid Atmospheric Processes With Good Performance at Reduced Precision},
journal = {Geophysical Research Letters},
volume = {48},
number = {6},
pages = {e2020GL091363},
doi = {10.1029/2020GL091363},
url = {https://agupubs.onlinelibrary.wiley.com/doi/abs/10.1029/2020GL091363},
note = {e2020GL091363 2020GL091363},
year = {2021},
}

@article{atkinson_ftorch_2025, 
doi = {10.21105/joss.07602}, 
url = {https://doi.org/10.21105/joss.07602}, 
year = {2025}, 
publisher = {The Open Journal}, 
volume = {10}, 
number = {107}, 
pages = {7602}, 
author = {Atkinson, Jack and Elafrou, Athena and Kasoar, Elliott and Wallwork, Joseph G. and Meltzer, Thomas and Clifford, Simon and Orchard, Dominic and Edsall, Chris}, 
title = {FTorch: a library for coupling PyTorch models to Fortran}, 
journal = {Journal of Open Source Software} 
}

@article{craig_development_2017,
	title = {Development and performance of a new version of the {OASIS} coupler, {OASIS3}-{MCT}\_3.0},
	volume = {10},
	url = {https://gmd.copernicus.org/articles/10/3297/2017/},
	doi = {10.5194/gmd-10-3297-2017},
	number = {9},
	journal = {Geoscientific Model Development},
	author = {Craig, A. and Valcke, S. and Coquart, L.},
	year = {2017},
	pages = {3297--3308},
}

@article{ukkonen_exploring_2022,
author = {Ukkonen, Peter},
title = {Exploring Pathways to More Accurate Machine Learning Emulation of Atmospheric Radiative Transfer},
journal = {Journal of Advances in Modeling Earth Systems},
volume = {14},
number = {4},
pages = {e2021MS002875},
doi = {10.1029/2021MS002875},
url = {https://agupubs.onlinelibrary.wiley.com/doi/abs/10.1029/2021MS002875},
note = {e2021MS002875 2021MS002875},
year = {2022}
}

@article{giorgetta_icon-A_2018,
author = {Giorgetta, M. A. and Brokopf, R. and Crueger, T. and Esch, M. and Fiedler, S. and Helmert, J. and Hohenegger, C. and Kornblueh, L. and Köhler, M. and Manzini, E. and Mauritsen, T. and Nam, C. and Raddatz, T. and Rast, S. and Reinert, D. and Sakradzija, M. and Schmidt, H. and Schneck, R. and Schnur, R. and Silvers, L. and Wan, H. and Zängl, G. and Stevens, B.},
title = {ICON-A, the Atmosphere Component of the ICON Earth System Model: I. Model Description},
journal = {Journal of Advances in Modeling Earth Systems},
volume = {10},
number = {7},
pages = {1613-1637},
doi = {10.1029/2017MS001242},
url = {https://agupubs.onlinelibrary.wiley.com/doi/abs/10.1029/2017MS001242},
year = {2018}
}

@article{yu_climSim-Online_2025,
  author  = {Sungduk Yu and Zeyuan Hu and Akshay Subramaniam and Walter Hannah and Liran Peng and Jerry Lin and Mohamed Aziz Bhouri and Ritwik Gupta and Bj{{\"o}}rn L{{\"u}}tjens and Justus C. Will and Gunnar Behrens and Julius J. M. Busecke and Nora Loose and Charles I Stern and Tom Beucler and Bryce Harrop and Helge Heuer and Benjamin R Hillman and Andrea Jenney and Nana Liu and Alistair White and Tian Zheng and Zhiming Kuang and Fiaz Ahmed and Elizabeth Barnes and Noah D. Brenowitz and Christopher Bretherton and Veronika Eyring and Savannah Ferretti and Nicholas Lutsko and Pierre Gentine and Stephan Mandt and J. David Neelin and Rose Yu and Laure Zanna and Nathan M. Urban and Janni Yuval and Ryan Abernathey and Pierre Baldi and Wayne Chuang and Yu Huang and Fernando Iglesias-Suarez and Sanket Jantre and Po-Lun Ma and Sara Shamekh and Guang Zhang and Michael Pritchard},
  title   = {ClimSim-Online: A Large Multi-Scale Dataset and Framework for Hybrid Physics-ML Climate Emulation},
  journal = {Journal of Machine Learning Research},
  year    = {2025},
  volume  = {26},
  number  = {142},
  pages   = {1--85},
  eprint  = {http://jmlr.org/papers/v26/24-1014.html}
}

@inproceedings{yu_climSim_2023,
title={ClimSim: A large multi-scale dataset for hybrid physics-{ML} climate emulation},
author={Sungduk Yu and Walter Hannah and Liran Peng and Jerry Lin and Mohamed Aziz Bhouri and Ritwik Gupta and Bj{\"o}rn L{\"u}tjens and Justus Christopher Will and Gunnar Behrens and Julius Busecke and Nora Loose and Charles I Stern and Tom Beucler and Bryce Harrop and Benjamin R Hillman and Andrea Jenney and Savannah Ferretti and Nana Liu and Anima Anandkumar and Noah D Brenowitz and Veronika Eyring and Nicholas Geneva and Pierre Gentine and Stephan Mandt and Jaideep Pathak and Akshay Subramaniam and Carl Vondrick and Rose Yu and Laure Zanna and Tian Zheng and Ryan Abernathey and Fiaz Ahmed and David C Bader and Pierre Baldi and Elizabeth Barnes and Christopher Bretherton and Peter Caldwell and Wayne Chuang and Yilun Han and YU HUANG and Fernando Iglesias-Suarez and Sanket Jantre and Karthik Kashinath and Marat Khairoutdinov and Thorsten Kurth and Nicholas Lutsko and Po-Lun Ma and Griffin Mooers and J. David Neelin and David Randall and Sara Shamekh and Mark A Taylor and Nathan Urban and Janni Yuval and Guang Zhang and Michael Pritchard},
booktitle={Thirty-seventh Conference on Neural Information Processing Systems Datasets and Benchmarks Track},
year={2023},
url={https://openreview.net/forum?id=W5If9P1xqO}
}

@article{chevallier_neural_1998,
  TITLE = {{A Neural Network Approach for a Fast and Accurate Computation of a Longwave Radiative Budget}},
  AUTHOR = {Chevallier, F. and Ch{\'e}ruy, F. and Scott, N. and Ch{\'e}din, A.},
  URL = {https://hal.science/hal-02946326},
  JOURNAL = {{Journal of Applied Meteorology}},
  PUBLISHER = {{American Meteorological Society}},
  VOLUME = {37},
  NUMBER = {11},
  PAGES = {1385-1397},
  YEAR = {1998},
  MONTH = Nov,
  doi =   {10.1175/1520-0450(1998)037%3C1385:ANNAFA%3E2.0.CO;2},
  HAL_ID = {hal-02946326},
  HAL_VERSION = {v1},
}

@article{krasnopolsky_new_2005,
author = {Krasnopolsky, Vladimir and Fox-Rabinovitz, Michael and Chalikov, Dmitry},
year = {2005},
month = {05},
pages = {1370-1383},
title = {New Approach to Calculation of Atmospheric Model Physics: Accurate and Fast Neural Network Emulation of Longwave Radiation in a Climate Model},
volume = {133},
journal = {Monthly Weather Review - MON WEATHER REV},
doi = {10.1175/MWR2923.1}
}

@article{brenowitz_prognostic_2018,
author = {Brenowitz, N. D. and Bretherton, C. S.},
title = {Prognostic Validation of a Neural Network Unified Physics Parameterization},
journal = {Geophysical Research Letters},
volume = {45},
number = {12},
pages = {6289-6298},
doi = {10.1029/2018GL078510},
url = {https://agupubs.onlinelibrary.wiley.com/doi/abs/10.1029/2018GL078510},
year = {2018}
}

@article{ogorman_2018,
	title = {Using {Machine} {Learning} to {Parameterize} {Moist} {Convection}: {Potential} for {Modeling} of {Climate}, {Climate} {Change}, and {Extreme} {Events}},
	volume = {10},
	copyright = {©2018. The Authors.},
	issn = {1942-2466},
	shorttitle = {Using {Machine} {Learning} to {Parameterize} {Moist} {Convection}},
	url = {https://onlinelibrary.wiley.com/doi/abs/10.1029/2018MS001351},
	doi = {10.1029/2018MS001351},
	language = {en},
	number = {10},
	urldate = {2025-07-16},
	journal = {Journal of Advances in Modeling Earth Systems},
	author = {O'Gorman, Paul A. and Dwyer, John G.},
	year = {2018},
	pages = {2548--2563},
}

@article{harris_scientific_2021,
    title={A Scientific Description of the GFDL Finite-Volume Cubed-Sphere Dynamical Core},
    author={Harris, Lucas and Chen, Xi and Putman, William and Zhou, Linjiong and Chen, Jan-Huey},
    year={2021},
    journal={NOAA technical memorandum OAR GFDL ; 2021-001},
    doi={10.25923/6nhs-5897}
}

@article {zhou_toward_2019,
      author = "Linjiong Zhou and Shian-Jiann Lin and Jan-Huey Chen and Lucas M. Harris and Xi Chen and Shannon L. Rees",
      title = "Toward Convective-Scale Prediction within the Next Generation Global Prediction System",
      journal = "Bulletin of the American Meteorological Society",
      year = "2019",
      publisher = "American Meteorological Society",
      address = "Boston MA, USA",
      volume = "100",
      number = "7",
      doi = "10.1175/BAMS-D-17-0246.1",
      pages=      "1225 - 1243",
      url = "https://journals.ametsoc.org/view/journals/bams/100/7/bams-d-17-0246.1.xml"
}

@article{rasch_e3sm_2019,
author = {Rasch, P. J. and Xie, S. and Ma, P.-L. and Lin, W. and Wang, H. and Tang, Q. and Burrows, S. M. and Caldwell, P. and Zhang, K. and Easter, R. C. and Cameron-Smith, P. and Singh, B. and Wan, H. and Golaz, J.-C. and Harrop, B. E. and Roesler, E. and Bacmeister, J. and Larson, V. E. and Evans, K. J. and Qian, Y. and Taylor, M. and Leung, L. R. and Zhang, Y. and Brent, L. and Branstetter, M. and Hannay, C. and Mahajan, S. and Mametjanov, A. and Neale, R. and Richter, J. H. and Yoon, J.-H. and Zender, C. S. and Bader, D. and Flanner, M. and Foucar, J. G. and Jacob, R. and Keen, N. and Klein, S. A. and Liu, X. and Salinger, A.G. and Shrivastava, M. and Yang, Y.},
title = {An Overview of the Atmospheric Component of the Energy Exascale Earth System Model},
journal = {Journal of Advances in Modeling Earth Systems},
volume = {11},
number = {8},
pages = {2377-2411},
doi = {10.1029/2019MS001629},
url = {https://agupubs.onlinelibrary.wiley.com/doi/abs/10.1029/2019MS001629},
year = {2019}
}

@article{alexeev_pytorch-fortran_2023,
    author = {Alexeev, Dmitry},
    year = {2023},
    title = {alexeedm/pytorch-fortran: Version v0.4 (v0.4)},
    journal = {zenodo},
    doi = {10.5281/zenodo.7851167}
}

@article {brenowitz_interpreting_2020,
      author = "Noah D. Brenowitz and Tom Beucler and Michael Pritchard and Christopher S. Bretherton",
      title = "Interpreting and Stabilizing Machine-Learning Parametrizations of Convection",
      journal = "Journal of the Atmospheric Sciences",
      year = "2020",
      publisher = "American Meteorological Society",
      address = "Boston MA, USA",
      volume = "77",
      number = "12",
      doi = "10.1175/JAS-D-20-0082.1",
      pages=      "4357 - 4375",
      url = "https://journals.ametsoc.org/view/journals/atsc/77/12/jas-d-20-0082.1.xml"
}

@article{bony_cfmip_2011,
    author = {S. Bony and M. Webb and C. Bretherton and S. Klein and P. Siebesma  and G. Tselioudis and M. Zhang},
    title = {CFMIP:
Towards a better evaluation and understanding of clouds and cloud feedbacks in CMIP5 models},
    journal = {CLIVAR Exchanges, Special Issue on the WCRP Coupled Model Intercomparison Project – Phase 5 (CMIP5)},
    number = {56},
    volume = {16},
    issue = {2},
    pages = {20 - 24},
    year = {2011},
    eprint = {https://www.clivar.org/sites/default/files/documents/Exchanges56.pdf},
}

@Article{steininger_density-based_2021,
author={Steininger, Michael
and Kobs, Konstantin
and Davidson, Padraig
and Krause, Anna
and Hotho, Andreas},
title={Density-based weighting for imbalanced regression},
journal={Machine Learning},
year={2021},
month={Aug},
day={01},
volume={110},
number={8},
pages={2187-2211},
issn={1573-0565},
doi={10.1007/s10994-021-06023-5},
url={https://doi.org/10.1007/s10994-021-06023-5},
}

@article{Beucler_2024,
    author = {Tom Beucler  and Pierre Gentine  and Janni Yuval  and Ankitesh Gupta  and Liran Peng  and Jerry Lin  and Sungduk Yu  and Stephan Rasp  and Fiaz Ahmed  and Paul A. O’Gorman  and J. David Neelin  and Nicholas J. Lutsko  and Michael Pritchard },
    title = {Climate-invariant machine learning},
    journal = {Science Advances},
    volume = {10},
    number = {6},
    pages = {eadj7250},
    year = {2024},
    doi = {10.1126/sciadv.adj7250},
    URL = {https://www.science.org/doi/abs/10.1126/sciadv.adj7250},
}

@article{Manabe1967,
      author = "Syukuro  Manabe and Richard T.  Wetherald",
      title = "Thermal Equilibrium of the Atmosphere with a Given Distribution of Relative Humidity",
      journal = "Journal of Atmospheric Sciences",
      year = "1967",
      publisher = "American Meteorological Society",
      address = "Boston MA, USA",
      volume = "24",
      number = "3",
      doi = "10.1175/1520-0469(1967)024%3C0241:TEOTAW%3E2.0.CO;2",
      pages=      "241 - 259",
      url = "https://journals.ametsoc.org/view/journals/atsc/24/3/1520-0469_1967_024_0241_teotaw_2_0_co_2.xml"
}

@article{xie_2017,
	title = {Reprocessed, {Bias}-{Corrected} {CMORPH} {Global} {High}-{Resolution} {Precipitation} {Estimates} from 1998},
	volume = {18},
	issn = {1525-7541, 1525-755X},
	url = {https://journals.ametsoc.org/view/journals/hydr/18/6/jhm-d-16-0168_1.xml},
	doi = {10.1175/JHM-D-16-0168.1},
	language = {EN},
	number = {6},
	journal = {Journal of Hydrometeorology},
	author = {Xie, Pingping and Joyce, Robert and Wu, Shaorong and Yoo, Soo-Hyun and Yarosh, Yelena and Sun, Fengying and Lin, Roger},
	month = jun,
	year = {2017},
	pages = {1617--1641},
}

@article{huffman_2019,
	title = {Integrated multi-satellite retrievals for {GPM} ({IMERG}) technical documentation. {NASA} {Tech} {Doc}., 77 pp},
	author = {Huffman, G. J. and Bolvin, D. T. and Nelkin, E. J. and Tan, J.},
    eprint = {https://gpm.nasa.gov/sites/default/files/document_files/IMERG_doc_190909.pdf},
    year = {2019},
}

@article{neale_cam5_2012, 
    title = {Description of the NCAR Community Atmosphere Model (CAM 5.0)},
    author = {Richard B. Neale and Andrew Gettelman and Sungsu Park and Chih-Chieh Chen and Peter H. Lauritzen and David L. Williamson and Andrew J. Conley and Doug Kinnison and Dan Marsh and Anne K. Smith and Francis Vitt and Rolando Garcia and Jean-Francois Lamarque and Mike Mills and Simone Tilmes and Hugh Morrison and Philip Cameron-Smith and William D. Collins and Michael J. Iacono and Richard C. Easter and Xiaohong Liu and Steven J. Ghan and Philip J. Rasch and Mark A. Taylor},
    eprint = {https://doi.org/10.5065/wgtk-4g06}, 
    year = {2012},
}

@article{geoffroy_global_2025,
Author = {Olivier Geoffroy and David Saint-Martin},
Title = {Global Kilometer-Scale Simulations with ARP-GEM2: Effect of Parameterized Convection and Calibration},
Year = {2025},
Eprint = {arXiv:2511.00829},
}

@article{nesbitt_storm_2006,
	title = {Storm {Morphology} and {Rainfall} {Characteristics} of {TRMM} {Precipitation} {Features}},
	volume = {134},
	issn = {1520-0493, 0027-0644},
	url = {https://journals.ametsoc.org/view/journals/mwre/134/10/mwr3200.1.xml},
	doi = {10.1175/MWR3200.1},
	number = {10},
	urldate = {2026-02-12},
	journal = {Monthly Weather Review},
	author = {Nesbitt, Stephen W. and Cifelli, Robert and Rutledge, Steven A.},
	month = oct,
	year = {2006},
	note = {Publisher: American Meteorological Society
Section: Monthly Weather Review},
	pages = {2702--2721},
}

@article{cess_methodology_1988,
	title = {A methodology for understanding and intercomparing atmospheric climate feedback processes in general circulation models},
	volume = {93},
	copyright = {Copyright 1988 by the American Geophysical Union.},
	issn = {2156-2202},
	doi = {10.1029/JD093iD07p08305},
	language = {en},
	number = {D7},
	urldate = {2026-02-19},
	journal = {Journal of Geophysical Research: Atmospheres},
	author = {Cess, Robert D. and Potter, Gerald L.},
	year = {1988},
	pages = {8305--8314},
}

\end{document}